\documentclass[12pt]{article}
\pdfoutput=1
\usepackage[a4paper]{geometry}
\usepackage{jheppub, amsmath,amssymb,amsfonts,amsxtra, mathrsfs, makeidx,graphics,graphicx,amsthm,epsfig, ytableau,bm, longtable,float, color,tikz,mathtools,xfrac,footnote,rotating, lscape}
\usetikzlibrary{positioning}
\usetikzlibrary{chains}
\usetikzlibrary{arrows,fit, decorations.pathreplacing}
\definecolor{darkgreen}{rgb}{0.0, 0.2, 0.13}

\usepackage{bbm}

\numberwithin{equation}{section}

\newcommand{\be}{\begin{equation}} \newcommand{\ee}{\end{equation}}
\newcommand{\bea}{\begin{equation} \begin{aligned}} \newcommand{\eea}{\end{aligned} \end{equation}}

\def\rt2{\sqrt{2}}

\def\mod{{\rm mod}}

\def\Tr{\mathop{\rm Tr}}

\def\CC{{\cal C}}

\def\CH{{\cal H}}

\def\CN{{\cal N}}

\def\1{{\ds 1}}

\def\repa{\raise4pt\hbox{$\square$}\mkern-14mu\raise-4pt\hbox{$\square$}}
\def\repab{\overline{\raise4pt\hbox{$\square$}\mkern-14mu\raise-4pt\hbox{$\square$}\mkern-1mu}}

\def\smileface{\ensuremath{\hbox{\large$\bigcirc$}\mkern-15mu\raise-1pt\hbox{\scriptsize$\smallsmile$}%
\mkern-10mu\raise4pt\hbox{..}\mkern4mu}}
\def\frownface{\ensuremath{\hbox{\large$\bigcirc$}\mkern-15mu\raise-1pt\hbox{\scriptsize$\smallfrown$}%
\mkern-10mu\raise4pt\hbox{..}\mkern4mu}}

\def\node#1#2{\overset{#1}{\underset{#2}{\circ}}}

\def\bnode#1#2{\overset{#1}{\underset{#2}{{\color{blue} \bullet}}}}
\def\gnode#1#2{\overset{#1}{\underset{#2}{{\color{gray} \bullet}}}}

\def\gver#1#2{\overset{{\llap{$\scriptstyle#1$}\displaystyle{\color{gray} \bullet}{\rlap{$\scriptstyle#2$}}}}{\scriptstyle\vert}}
\def\ver#1#2{\overset{{\llap{$\scriptstyle#1$}\displaystyle\circ{\rlap{$\scriptstyle#2$}}}}{\scriptstyle\vert}}
\def\wver#1#2{\overset{{\llap{$\scriptstyle#1$}\displaystyle{\square}{\rlap{$\scriptstyle#2$}}}}{\scriptstyle\vert}}

\def\redsqver#1#2{\overset{{\llap{$\scriptstyle#1$}\displaystyle{\color{red} \blacksquare}{\rlap{$\scriptstyle#2$}}}}{\scriptstyle\vert}}

\tikzstyle{every picture}+=[remember picture]
\tikzstyle{na} = [baseline=-.5ex]

\newtheorem{conj}{Conjecture}

\newcommand{\ba}{\begin{array}}
\newcommand{\ea}{\end{array}}
\newcommand{\bi}{\begin{itemize}}
\newcommand{\ei}{\end{itemize}}

\def\bea#1\eea{\allowdisplaybreaks \begin{align}#1\end{align}}
 \newcommand{\ben}{\begin{enumerate}}
\newcommand{\een}{\end{enumerate}}
\newcommand{\bean}{\begin{eqnarray*}}
\newcommand{\eean}{\end{eqnarray*}}
\newcommand{\eref}[1]{(\ref{#1})}

\newcommand{\PE}{\mathop{\rm PE}}

\newcommand{\BC}{\mathbb{C}}

\newcommand{\BZ}{\mathbb{Z}}

\newcommand{\BH}{\mathbb{H}}

\newcommand{\comment}[1]{}

\newcommand{\red}{\color{red}}

\title{Discrete Gauging in Six Dimensions}
\author[a]{Amihay Hanany}
\author[b]{, and Gabi Zafrir}

\affiliation[a]{Theoretical Physics Group, Imperial College London, \\
	Prince Consort Road, London, SW7 2AZ, UK}
\affiliation[b]{Kavli Institute for the Physics and Mathematics of the Universe, \\ University of Tokyo, Kashiwa, Chiba 277-8583, Japan}

\emailAdd{a.hanany@imperial.ac.uk}
\emailAdd{gabi.zafrir@ipmu.jp}

\preprint{
	{\small
		\begin{flushright}
			IMPERIAL-TP-18-AH-03\\
		\end{flushright}
	}
}

\abstract{When $n$ M5 branes coincide on an A type singularity, $\BC^2/\BZ_k$, there is a multitude of tensionless strings which arise in the spectrum. The low energy theory when all M5 branes are separated at the singularity is given by a linear quiver with parameters $n$ and $k$. The theory has a multitude of phases, as many as partitions of $n$, each characterized by a different Higgs branch. Each such Higgs branch can be described by a Coulomb branch of a 3d $\CN=4$ quiver. For example, at finite coupling, when all branes are separated, the quiver has a bouquet of $n$ $U(1)$ nodes connected to a single node. There is a natural discrete non Abelian $S_n$ global symmetry which acts in the theory by permuting $n$ identical objects. It acts in particular on the Higgs branch at the above finite coupling phase. It is conjectured that at the coincident point this discrete $S_n$ flavor symmetry is gauged, and at partial coincidence the corresponding subgroup of $S_n$ is gauged. This elegant and simple effect solves several problems which are raised recently on the physics of multiple M5 branes on an A type singularity. Similar results on multitude of phases are concluded for a system of $n$ M5 branes on an A type singularity next to an M9 plane.
}
\begin{document}
\maketitle

\section{Introduction and Summary}
A particularly simple class of 6d $\CN=(1,0)$ supersymmetric theories is given by a set of $n$ M5 branes on an ALE singularity, $\BC^2/\Gamma$ where $\Gamma\subset SU(2)$ is a discrete subgroup of either cyclic, binary dihedral, or (binary) tetrahedral, icosahedral or dodecahedral type. Such theories are discussed recently in \cite{DelZotto:2014hpa} where it is pointed out that at the fixed point there are 0, 1, 3, 5, and 11 new massless tensor multiplets, respectively, in addition to vector multiplets, hyper multiplets, and other exotic matter termed conformal. The 6d theory has no Lagrangian, which motivates the study of such theories using other techniques. A crucial role for understanding the physics of such theories is played by the elliptic genus which is studied in a series of papers such as \cite{Haghighat:2013tka} and works which follow.

The approach taken in a recent set of papers \cite{Cremonesi:2015lsa, Ferlito:2017xdq, Hanany:2018uhm}, and in the present paper, is to study the behavior of the Higgs branch as one tunes a gauge coupling to infinity, or in 4 dimensions on an Argyres Douglas point at finite coupling \cite{DelZotto:2014kka}. Such an approach turns out to be particularly useful in revealing new physics related to instanton operators \cite{Cremonesi:2015lsa, Ferlito:2017xdq}, small $E_8$ instanton transitions \cite{Ganor:1996mu}, Kraft - Procesi (KP) transitions \cite{Cabrera:2016vvv, Cabrera:2017njm, Cabrera:2018ann} and the physics of a single M5 brane on a D type singularity \cite{Hanany:2018uhm}. A key feature in this approach is to express the Higgs branch at infinite coupling as a Coulomb branch of a 3d $\CN=4$ quiver gauge theory. The techniques of \cite{Cremonesi:2013lqa} prove to be very useful in evaluating the Coulomb branch of such quivers and as a result the Higgs branch of the 6d SCFT is evaluated.

This paper is devoted to the study of $n$ M5 branes on a $\BC^2/\BZ_k$ singularity. A crucial difference between this case and the other D and E singularities is that there are no additional massless tensor multiplets at the limit of zero tension, and hence there is no phenomenon which is related to small instanton transitions. Instead there are few recent observations about the global symmetry at the coincident point, summarized in \cite{Bah:2017gph}, which we now review.

For general values of $n$ and $k$ the global symmetry is $S(U(k)\times U(k))$. For $k = 2$ this global symmetry is enhanced to $SU(2)^3$. For $n = 2$ the global symmetry is enhanced to $SU(2k)$. For $k = n = 2$ the global symmetry is enhanced to $SO(7)$. For $n = 1$, we have $k^2$ six dimensional free hypermultiplets which transform in the bifundamental representation of the global symmetry. These global symmetries should be represented by isometries of the Higgs branch of the SCFT. On the other hand, at the separated case, where all gauge couplings are finite, the Higgs branch generally develops additional isometries. For general values of $n$ and $k$ it has an $S(U(k)\times U(k)\times U(1)^{n-1})$ isometry, which for $k = 2$ is enhanced to $SU(2)^{n+2}$. For $n = 2$ it is enhanced to $S(U(2k)\times U(1))$. For $k = n = 2$ it is enhanced to $SO(8)$. It should be noted that most $U(1)$ isometries are not global symmetries as these are anomalous in 6 dimensions. Nevertheless one can still use grading with respect to such $U(1)$ isometries along the Higgs branch, since these are not anomalous in lower dimensions, while the Higgs branch remains the same. Most such $U(1)$ isometries are lost at the tensionless limit.

These isometries, realized on the Higgs branch, pose a challenge of explaining the pattern of breaking, as in the case $k=n=2$ where the $SO(8)$ symmetry at finite coupling is broken to $SO(7)$ at infinite coupling, or explaining the pattern of loss of $U(1)$ isometries on the Higgs branch, as in the case of $n=2$ where the Higgs branch has a $U(1)$ isometry at finite coupling but loses it at infinite coupling.

In addition to the continuous global symmetries, there are additional discrete global symmetries, rarely discussed in the literature, which are going to play a very crucial role in what follows. For the theory at finite coupling, where all M5 branes are separated, there is a discrete $S_n$ global symmetry, where $S_n$ is the permutation group in $n$ elements. This $S_n$ acts on various observables such as the $n$ $U(1)$ global symmetries at finite coupling. It also acts on operators in the ring of BPS operators on the Higgs branch, producing an interesting pattern of transformation laws under this discrete symmetry $S_n$ and its representation theory.

An elegant description that explains all the patterns in the behavior of the isometries, and more generally the behavior of the fixed point theory and its Higgs branch in particular, is the notion of discrete gauging in six dimensions. In this paper we propose that {\it as $n$ M5 branes coincide, the $S_n$ global symmetry is gauged}. This is the main result of the paper and the remaining part of the paper is devoted to the study of such a proposal, providing evidence and consequences of such a phenomenon.

Contrary to the case of an M5 brane on a D type singularity, there is no small instanton transition and hence the number of hyper multiplets does not change as one tunes the inverse gauge coupling to 0.
For the D type singularity the dimension of the Higgs branch jumps as one goes to the tensionless limit, while for A type singularity we do not expect the dimension of the Higgs branch to change. Indeed, this is consistent with discrete gauging, as the dimension of the moduli space does not change, while the operator content gets reduced to the invariant sector under $S_n$. This is going to serve as a crucial test for our proposal. So we learn that while for the D type singularity there are extra massless states from the limit of zero tension, for the A type singularity there is a loss of states at the tensionless limit, though without losing flat directions.

The low energy theory of $n$ separated M5 branes has a collection of $n-1$ massless tensor multipets (excluding the central decoupled tensor multiplet), with a real scalar in each tensor multiplet. These scalars serve as tensions of BPS strings and/or as gauge couplings for gauge fields. Each time one or more of the BPS tensions goes to 0, and/or one of the inverse gauge couplings are tuned to 0, we get a new phase which, as discussed below, is characterized by a different Higgs branch. As there are as many phases as there are partitions of $n$, we expect to find such a number of different Higgs branches. As mentioned above, the best way to describe these different Higgs branches is using a Coulomb branch of a 3d $\CN=4$ quiver, and one can view the Higgs branch as an object, like an order parameter, that characterizes the phase of the theory.

Geometrically, the phenomenon of discrete gauging brings an interesting relation between hyper K\"ahler cones. Given a space with a discrete isometry $S_n$ one can construct a new space by gauging $H$ a discrete subgroup of $S_n$. For $\lambda$ a given partition of $n$ there is a corresponding subgroup $H_\lambda$ which physically corresponds to subsets of coincident branes as prescribed by the partition. This corresponds to partial cases where strings become tensionless, while other gauge couplings remain finite. For such a partial case one gets a discrete gauging of $H_\lambda$ which presents a rich pattern of moduli spaces, all descend from a parent theory, the Higgs branch at finite coupling with a global $S_n$ symmetry.

The results of this paper demonstrate that the Higgs branch of the six dimensional theory undergoes a discrete quotient by $S_n$ when $n$ M5 branes coincide. Since the physics of the problem is that of $n$ identical objects, it is natural to make a stronger conjecture that the global $S_n$ symmetry is gauged not just on the Higgs branch, but also on any observable or any physical quantity in the theory. In some cases the observables are singlets of $S_n$ and survive the projection, while other transform in irreducible representations of this global symmetry. The strong form of the conjecture is that the $S_n$ symmetry is gauged in all sectors. As the evidence of this paper deals with Higgs branches only, it remains as a crucial challenge to check the conjecture further in its stronger form.

%Let us begin by describing the gauge theory at the generic situation where all gauge couplings are finite and discuss some of its features.

In this paper we frequently deal with the Coulomb branches of 3d gauge theories. All the 3d Coulomb branches in this paper have the property that all nodes are gauge nodes of unitary type (not special unitary in particular), and all edges are of bi fundamental type, including adjoint matter between a node and itself. It excludes bi fundamental matter of non simply laced type, which do appear in various other moduli spaces. As a result, all 3d quivers have a Lagrangian, and there is an overall $U(1)$ gauge symmetry which can decouple from the theory. Also in actual computations one needs to pick a $U(1)$ factor and sets its magnetic charge to some constant value, The choice of such $U(1)$ is done upon convenience of computation, and is not affecting the final result. We comment that there is such a step in the computation, referring the details to \cite{Cremonesi:2013lqa} and papers which follow.

The structure of this article is as follows. In Section \eref{sectionM5} we consider the case of the 6d supersymmetric field theory (SFT), with or without tensionless strings, living on $n$ M5-branes on a $\BC^2/Z_k$ singularity. We state our conjecture for this case and perform several tests on it. Section \eref{Smallinstantons} examines the case of the 6d SFT living on $n$ M5-branes on a $\BC^2/Z_k$ singularity near an M9-plane in two examples. Finally Section \eref{relationship} provides motivations for the various 3d quivers, used to study the Higgs branches of the 6d theories at diverse strong coupling points, using recent, and not so recent, results on the compactification of 6d theories to 4d and 3d. 

\section{$n$ M5-branes on a $\BC^2/Z_k$ singularity}
\label{sectionM5}
The world volume theory of $n$ separated M5 branes on a $\BC^2/Z_k$ singularity is a 6d $\CN=(1,0)$ theory with $n-1$ tensor multiplets, a gauge group $SU(k)^{n-1}$ and bifundamental matter given by a linear quiver \cite{Intriligator:1997kq,  Brunner:1997gk,Blum:1997mm, Intriligator:1997dh, Brunner:1997gf, Hanany:1997gh, DelZotto:2014hpa}.
\be \label{linearquiver}
{\sf Q}_{n,k} = ~~\underbrace{\node{\wver{}{\,\,k}}{k} - \node{}{k} - \cdots - \node{}{k} - \node{\wver{}{\,\,k}}{k} 
}_{n-1~\text{\comment{$(k)$} nodes}}~, 
\ee
where each gauge node represents an $SU(k)$ gauge symmetry. The brane system can be depicted by the following Type IIA diagram (see for example \cite{Hanany:1997gh}) where D6 branes span one extra direction, called 6 in the diagram, and share the remaining 5+1 directions with the NS5 branes.

\be \label{brane}
\begin{tikzpicture}[baseline]
% NS5
\filldraw (0,1.24) circle (0.1cm) node[xshift =0cm, yshift=0.4cm] {\tiny NS5};
\filldraw (2,1.24) circle (0.1cm) node[xshift =0cm, yshift=0.4cm] {\tiny NS5};
% D6
\draw [very thick] (0.1,1.24)--(1.9,1.24) node[black,midway, yshift=-0.3cm] {\scriptsize $k$ D6};
\draw [very thick] (1.9,1.24)--(4,1.24) node[black,midway, yshift=-0.3cm] {\scriptsize $k$ D6};
\draw [very thick] (-2,1.24)--(0,1.24) node[black,midway, yshift=-0.3cm] {\scriptsize $k$ D6};
% O6
%\draw [darkgreen, dashed, very thick] (-2,1.28)--(4,1.28) node[black,xshift=0.5cm, yshift=0cm] 
{\scriptsize {\color{darkgreen} O6}} node[black,xshift=-5cm, yshift=0.2cm]
{\scriptsize \color{darkgreen} $ $} node[black,xshift=-3cm, yshift=0.2cm] {\scriptsize\color{darkgreen} $ $} node[black,xshift=-1cm, yshift=0.2cm] {\scriptsize \color{darkgreen} $ $}
node[black,xshift=2cm, yshift=0cm] {\scriptsize $\overset{x^6}{\longrightarrow}$};
% D2
\draw [red, very thick] (-0.5,1.28)--(1.4,1.28) node[black,xshift=-1.3cm, yshift=0.2cm] {\tiny \red D2};
\end{tikzpicture} 
\ee
This Type IIA diagram is helpful in identifying the low energy quiver theory that lives on the M5 branes, as the M theory dual turns the NS5 into M5 and the D6 into an A type singularity, while keeping the quiver description fixed.

Each of the $n-1$ tensor multiplets contains a real scalar field which serves as an inverse gauge coupling to each of the $n-1$ gauge fields and is given by the separation of two neighboring NS5 branes in the figure.  When two or more M5 branes coincide, we get BPS tensionless strings, depicted in the figure by the red D2 brane in Type IIA, or by an M2 brane in M theory, and the theory becomes strongly coupled. The main tool that we use in this paper is to ask what happens to the Higgs branch as one or more of the gauge couplings is tuned to infinity. There is naively an $S(U(k)\times U(k)\times U(1)^{n-1})$ global symmetry coming from the two flavor nodes at the ends and each node contributes a baryon number, though $n-1$ of the $U(1)$ symmetries are anomalous in six dimensions, and so do not contribute to the global symmetry. Nevertheless, they still contribute grading on functions on the Higgs branch. From the form of the quiver in \eref{linearquiver} it is not evident that there is a discrete $S_n$ global symmetry (one expects it as the Weyl group of the algebra given by the Dynkin diagram formed by the subset of balanced nodes). Instead, we see it in the 3d $\CN=4$ Coulomb branch description of the moduli space. Nevertheless, one should remember that the brane system is that of $n$ identical objects, and hence the discrete symmetry must be a feature of any low energy description of the separated M5 brane system.

For the special case of $n=2$ the quiver becomes
\be
{\sf Q}_{2,k} = ~~\node{\wver{}{\,\,2k}}{k} 
\ee
with a naive global symmetry $U(2k)$, where again the $U(1)$ global symmetry is anomalous in six dimensions.

The finite coupling Higgs branch $\CH_f$ of this theory has dimension $k^2+n-1$ where the linear dependence on $n$ indicates that we add one more baryonic direction to the moduli space for each M5 brane which is added to the system.
For the special case of $n=2$ we can record the highest weight generating function (HWG) \footnote{Recall from \cite{Hanany:2014dia} that the HWG encodes the representation content of all protected BPS operators in the ring of holomorphic operators on the Higgs branch.} \cite{Hanany:2014dia} for the Higgs branch at finite coupling $\CH_f$ by taking the result from Equation (4.10) of \cite{Benvenuti:2010pq} and generalizing to
\be\label{HWGf2k}
HWG(\CH_f)_{n=2, k} = \PE \left[\sum_{i=1}^{k-1} \mu_i \mu_{2k-i} t^{2i} +t^2 + \mu_k (q^k + q^{-k})t^k \right],
\ee
where $t$ is the fugacity for the $SU(2)_R$ symmetry on the Higgs branch, $\mu_i$ are the fugacities for highest weights of representations of $SU(2k)$ and $q$ is the fugacity for the $U(1)$ anomalous baryonic symmetry. The baryon number is normalized such that a quark has charge 1 and a baryon has charge $k$. The form of \eref{HWGf2k} is used below.

To proceed with a manifest $S_n$ global symmetry we now turn to the 3d $\CN=4$ description of this moduli space. To compute the corresponding quiver one may use two steps. In the first step study the theory where all gauge nodes are $U(k)$ and in the second step turn off all baryonic gauge couplings. The 3d mirror of quiver \eref{linearquiver} with all gauge nodes $U(k)$ is easy to compute by observing that the global symmetry on the Higgs branch contains two $SU(k)$ factors, and one $U(1)$ factor, as well as $n-1$ balanced nodes which imply an $SU(n)$ global symmetry on the 3d Coulomb branch of this quiver. Alternatively, it can be derived using S-duality on the Type IIB brane system which constructs \cite{Hanany:1996ie} the 3d quiver \eref{linearquiver}. Either way, we get the 3d mirror quiver
\be \label{mirrorUk}
\node{}{1}- \node{}{2}-\cdots -\node{}{k-1}-\gnode{\wver{}{\,\,n}}{k}-\node{}{k-1}-\cdots - \node{}{2}-\node{}{1}~, 
\ee
where the grey node indicates an unbalanced node, and all other are balanced.
The next step is to turn off all baryonic couplings and this results in replacing the flavor node of $n$ by $n$ gauge nodes of 1, a process which is termed implosion as in \cite{2012arXiv1209.1578D}\footnote{In the physics literature this expresses the fact that one can in a sense invert a $U(1)$ gauging by gauging the topological $U(1)$ associated with that symmetry, see \cite{Hanany:1996ie, KapStr,Witten:2003sl}.}. The resulting quiver has a bouquet of	 $n$ nodes of 1 which are all connected to the central node $k$,
%\be \label{mirrorSUk}
%{\sf F}_{n,k} =~~\node{}{1}- \node{}{2}-\cdots -\node{}{k-1}-\node{\overbrace{\verdiagdown{1}{}\cdots\verdiagup{}{\,\,1}}^{n~\text{nodes}}}{k}-\node{}{k-1}-\cdots - \node{}{2}-\node{}{1}~
%\ee

\be \label{mirrorSUk}
{\sf F}_{n,k} =~~\node{}{1}- \node{}{2} -\cdots -\node{}{k-1}-\gnode{\overbrace{{\overset{1\gnode{}{}\quad \ldots\quad\gnode{}{}1}{\setminus\cdots /}}}^{n~\text{nodes}}}{k}-\node{}{k-1}-\cdots - \node{}{2}-\node{}{1}~
\ee

%\be \label{mirrorSUk}
%{\sf F}_{n,k} =~~\node{}{1}- \node{}{2} -\cdots -\node{}{k-1}-\gnode{\overbrace{{\overset{1\gnode{}{} \ldots\gnode{}{}1}{\setminus /}}}^{n~\text{nodes}}}{k}-\node{}{k-1}-\cdots - \node{}{2}-\node{}{1}~
%\ee

This form of the quiver makes it manifest that the global symmetry includes a discrete factor $S_n$, as well as the continuous $SU(k)\times SU(k)\times U(1)^n$. We now turn to the Higgs branch at infinite coupling, where all string tensions are set to 0, and record the following conjecture.

\begin{conj}
\label{Ohmori}
At infinite coupling
%there is a recent conjecture \cite{Mekareeya:2017jgc} which proposes that
the quiver for $n$ coincident M5 branes on a $\BC^2/\BZ_k$ singularity takes the form
%\be \label{mirrorUk}
%{\sf I}_{n,k} = ~ ~ \node{}{1}- \node{}{2}-\cdots -\node{}{k-1}-\node{\ver{\subset}{\,\,n}}{k}-\node{}{k-1}-\cdots - \node{}{2}-\node{}{1}~, 
%\ee

\be \label{mirrorUkinfinite}
{\sf I}_{n,k} = ~ ~ \node{}{1}- \node{}{2}-\cdots -\node{}{k-1}-\node{\overset{\cap}{\ver{}{\,\,n}}}{k}-\node{}{k-1}-\cdots - \node{}{2}-\node{}{1}~, 
\ee
where the symbol next to the node $n$ indicates that there is an additional adjoint hyper multiplet under the $U(n)$ gauge symmetry.
\end{conj}
Arguments for this conjecture are given in Section \eref{relationship}.
Henceforth a node with adjoint matter coupled to it is called ``an adjoint $n$ node".
According to this conjecture, with several low rank hints given in \cite{Mekareeya:2017jgc} and a D type analog given in \cite{Hanany:2018uhm}, the Coulomb branch of the 3d $\CN=4$ quiver ${\sf I}_{n,k}$ is the Higgs branch at infinite coupling (that is when all inverse couplings are set to 0) of the quiver ${\sf Q}_{n,k}$,
\be
\CH_\infty \left( {\sf Q}_{n,k} \right) = \CC^{3d}\left( {\sf I}_{n,k} \right) .
\ee
We can now state the main conjecture of this paper.
\be
\label{3dgauging}
\CC^{3d}\left( {\sf I}_{n,k} \right) = \CC^{3d} \left( {\sf F}_{n,k} \right) / S_n
\ee
which is a statement on three dimensional quivers, without any reference to six dimensional physics. Some tests for this conjecture are provided in a companion paper \cite{HZ}, where the statement is more general than in this paper, changing any $m$ bouquet by an adjoint $m$ node resulting in $S_m$ gauging. Some examples are discussed below, but relation \eref{3dgauging} allows us to rewrite it for six dimensional theories
\begin{conj}
\label{discrete}
When $n$ M5 branes coincide on a $\BC^2/\BZ_k$ singularity, the discrete $S_n$ global symmetry is gauged, and the Higgs branch at the coincident point is an $S_n$ orbifold of $\CH_f$, the Higgs branch when all branes are separated.
\be
\CH_\infty \left( {\sf Q}_{n,k} \right) = \CH_f \left( {\sf Q}_{n,k} \right) / S_n .
\ee
\end{conj}
In other words, we conjecture that when $n$ M5 branes coincide on a $\BC^2/\BZ_k$ singularity, there is a gauging of the discrete global $S_n$ symmetry which acts on the M5 branes.

This result can be generalized as follows.
For a given partition $\{n_i\}$ of $n$ such that $\sum_i n_i = n$ the branes can coincide in a pattern of $n_i$ coincident branes at position $x_i$ along the 6 direction where all $x_i$ are different. The case where all branes are separated, denoted above by $f$, is denoted by the partition $\{1^n\}$, while the case where all branes are coincident, denoted above by $\infty$, is denoted by the partition $\{n\}$. For a general partition we get a partial configuration where not all M5 branes are coincident, but rather divide into smaller groups of coincident M5 branes. In such a case the discrete group which is gauged is $\prod_i S_{n_i}\subset S_n$ and the corresponding Higgs branch is given by the  following orbifold conjecture
\begin{conj}\label{partialcoincidence}
\be
\CH_{\{n_i\}} \left( {\sf Q}_{n,k} \right) = \CH_{\{1^n\}} \left( {\sf Q}_{n,k} \right) / \prod_i S_{n_i} .
\ee
The corresponding quiver is
\be \label{quiverpartitions}
{\sf F}_{\{n_i\},k} =~~\node{}{1}- \node{}{2} -\cdots -\node{}{k-1}-\node{\overbrace{{\overset{n_1~\overset{\cap}{\node{}{}}\quad \ldots\quad\overset{\cap}{\node{}{}}~n_l}{\setminus\cdots /}}}^{l~\text{nodes}}}{k}-\node{}{k-1}-\cdots - \node{}{2}-\node{}{1}~,
\ee
with the relation
\be
\CH_{\{n_i\}} \left( {\sf Q}_{n,k} \right) = \CC\left({\sf F}_{\{n_i\},k}\right).
\ee
\end{conj}

This is a generalization of Conjectures \eref{Ohmori} and \eref{discrete}, extending the form of the quiver to include a collection of adjoint nodes, as well as a multitude of strongly coupled phases -- as many as partitions of $n$. It is in this sense that the Higgs branch characterizes the different phases of the theory, as it looks different in each phase. Arguments in favor of this conjecture can again be found in Section \eref{relationship}.

For a use below we now state the main result of \cite{HZ}.
\begin{conj}
\label{Anton}
Given a 3d $\CN=4$ quiver ${\sf Q}_{\{1^n\}}$ with $n$ nodes of 1 attached to another node, say $k$, (gauge node or global node)
\be
{\sf Q}_{\{1^n\}} =~~\cdots ~ - \node{\overbrace{{\overset{1~{\node{}{}}\quad \ldots\quad{\node{}{}}~1}{\setminus\cdots /}}}^{n~\text{nodes}}}{k} - ~ \cdots ~,
\ee
one can construct a new 3d $\CN=4$ quiver ${\sf Q}_{\{n\}}$ with an adjoint $n$ node attached to $k$,
\be
{\sf Q}_{\{n\}} =~~\cdots ~ -\node{\overset{\cap}{\ver{}{\,\,n}}}{k} - ~ \cdots ~.
\ee
Then the following relation between the Coulomb branches of these quivers holds
\be
\CC \left ( {\sf Q}_{\{n\}} \right ) = \CC \left({\sf Q}_{\{1^n\}}  \right ) / S_n ~.
\ee
\end{conj}
For the special case where $k$ is a global node, we recover the expected symmetric product as demonstrated in Section 4 of \cite{Cremonesi:2013lqa}.

We now turn to some examples and some special cases.

\subsection{2 M5 branes on an $A_1$ singularity} 

This is the case $n=k=2$ where the gauge theory for separated M5 branes is $SU(2)$ with 4 flavors.
It can be described by the quiver
%\be
%\bnode{\rver{}{SO(8)}}{Sp(1)}
%\ee

\be
{\sf Q}_{2,2} = ~\overset {\overset {SO(8)}{{\color{red} \blacksquare}}} {\bnode {\vert}{Sp(1)}}
\ee

where for convenience the $Sp$ and $SO$ nodes are depicted in colors.
This theory has an $SO(8)$ global symmetry and has a HWG which takes the form
\be
\label{HWGf22}
HWG(\CH_f)_{n=2, k=2} = \PE \left[\mu_2 t^2 \right]
\ee
where $\mu_2$ is the fugacity for the highest root of $SO(8)$.
Since $\CH_f$ is a closure of a nilpotent orbit, we can easily write it as an algebraic variety.
\be
\CH_{f, n=2, k=2} = \overline{{\rm min}_{SO(8)}} = \left \{ M_{8\times 8} | M = - M^T, M^2 = 0, rank(M) \le 2\right\} .
\ee
of quaternionic dimension 5. This moduli space is the closure of the minimal nilpotent orbit of $SO(8)$.

The global symmetry for the case of 2 coincident M5 branes is reported to be $SO(7)$ \cite{Ohmori:2015pia} and the Higgs branch still has quaternionic dimension 5. There is only one nilpotent orbit of $SO(7)$ with this dimension. It is the closure of the next to minimal orbit. As an algebraic variety it takes the form
\be
\label{Hinf22}
\CH_{\infty, n=2, k=2} = \overline{{\rm n.min}_{SO(7)}} = \left \{ M_{7\times 7} | M = - M^T, M^3 = 0, \Tr(M^2) = 0, rank(M) \le 2\right\} .
\ee
It is natural to assume that this is the Higgs branch of the theory with 2 coincident M5 branes on a $\BC^2/\BZ_2$ singularity, as other hyper K\"ahler cones with an $SO(7)$ isometry are likely to have higher dimension. This argument based on dimension is likely to be correct, but we need a better argument. Indeed, a direct computation of the Coulomb branch of the quiver in \eref{mirrorUkinfinite} for the case of $n=k=2$,
%\be \label{infinite22}
%{\sf I}_{2,2} = ~ ~ \node{}{1}-\node{\ver{\subset}{\,\,2}}{2}-\node{}{1}~, 
%\ee

\be \label{infinite22}
{\sf I}_{2,2} = ~ ~ \node{}{1}-\node{\overset{\cap}{\ver{}{\,\,2}}}{2}-\node{}{1}~, 
\ee
reveals that it is indeed correct, verifying equation \eref{Hinf22}.
Furthermore, by a classic result of Kostant and Brylinski \cite{1992math......4227B} this nilpotent orbit is the $\BZ_2$ quotient of the closure of the minimal nilpotent orbit of $SO(8)$. In other words, we confirm that
\be
\CH_\infty \left( {\sf Q}_{2,2} \right) = \CH_f \left( {\sf Q}_{2,2} \right) / S_2 .
\ee
This is the simplest test of conjecture \eref{discrete}.
The HWG for the moduli space $\overline{{\rm n.min}_{SO(7)}}$ was computed in \cite{Hanany:2016gbz} and was found to be
\be
\label{HWGi22}
HWG(\CH_\infty)_{n=2, k=2} = \PE \left[\mu_2 t^2 + \mu_1^2 t^4\right],
\ee
where $\mu_i$ are fugacities for highest weights of $SO(7)$.
By comparing \eref{HWGf22} with \eref{HWGi22} we see how the $\BZ_2$ acts on the different representations of $SO(8)$. The adjoint representation of $SO(8)$ decomposes to the adjoint and vector representations of $SO(7)$. In highest weight fugacities we can write the relation
\be
\label{branch87}
\mu_2 t^2 \longrightarrow (\mu_2 + \mu_1)t^2 .
\ee
The adjoint representation of $SO(7)$ is invariant under this $\BZ_2$ action, while the vector representation is in the non trivial representation. The natural invariant is then $\mu_1^2 t^4$ and \eref{HWGi22} is derived from \eref{HWGf22}.
This computation helps derive new results which are presented below.

Of course, one can use the Molien invariant (see for example a presentation in \cite{Benvenuti:2006qr}) and rewrite \eref{HWGi22} in the form
\be
\label{molien22}
HWG(\CH_\infty)_{n=2, k=2} = \frac{1}{1-\mu_2t^2} \frac{1}{2} \left( \frac{1}{1-\mu_1 t^2} + \frac{1}{1+\mu_1t^2} \right),
\ee
where the term in front of the bracket represents the invariant sector, while the first and second terms in the bracket are the contributions of the trivial and non trivial representations of $\BZ_2$, respectively.

\subsection{2 M5 branes on $\BC^2/\BZ_k$ singularity}
Given the success of the case $n=k=2$ we now turn to the general $k$ case. First let us compute the HWG for the Higgs branch of the theory at infinite coupling by applying a $\BZ_2$ quotient of the HWG in \eref{HWGf2k}. We then confirm that this is the correct result by direct computation. To figure out the action of $\BZ_2$ we revisit the case $k=2$ and notice that the global symmetry of \eref{HWGf2k} when $k=2$ is actually $SU(4)\cong SO(6)$ (here we are using a sloppy notation which does not make a distinction between $SO$ and $Spin$, as it is not relevant.) So we first decompose the $SO(8)$ representations into $SO(6)$
representations. The adjoint representation of $SO(8)$ decomposes to the adjoint, 2 vectors and one singlet of $SO(6)$ which can be summarized by
\be
\label{branch86}
\mu_2 t^2 \longrightarrow \left(\mu_1\mu_3 + \mu_2\left(q^2 + q^{-2}\right) + 1\right)t^2 .
\ee
where $\mu_1, \mu_2, \mu_3$ are the fugacities for highest weights of the fundamental, second rank antisymmetric, and anti fundamental representations of $SU(4)$, respectively. $q$ is a fugacity for a $U(1)$ which commutes with $SU(4)$ inside $SO(8)$, but it is not going to play a crucial role, so we set it to 1 below.
Comparing with \eref{branch87} we find that one vector and the singlet are in the nontrivial representation of $\BZ_2$ while the adjoint and the other vector are invariant. Since two terms are nontrivial, there are 3 invariants which satisfy a condition. They are $\mu_2^2 t^4$, $\mu_2 t^4$ and $t^4$, with a condition $\mu_2^2t^8$. Putting this together we get
\be
\label{HWG226}
HWG(\CH_\infty)_{n=2, k=2} = \PE \left[\left (\mu_1\mu_3 +\mu_2 \right )t^2 + \left (\mu_2^2 + \mu_2 + 1\right ) t^4 - \mu_2^2t^8 \right].
\ee
Where we recall that the $\mu$ are fugacities of $SU(4)$. One can check that the expressions in \eref{HWG226} and \eref{HWGi22} agree term by term in powers of t, except that in \eref{HWG226} we use representations of $SU(4)$ and in \eref{HWGi22} we use representations of $SO(7)$. This form is now ready to be generalized to any value of $k$, but let us quote the form of the Molien sum for expression \eref{HWG226} as well

\be
\label{HWG226molien}
HWG(\CH_\infty)_{n=2, k=2} = \frac{1}{2(1-\mu_1\mu_3t^2)(1-\mu_2t^2)} \left(\frac{1}{(1-t^2)(1-\mu_2t^2)} + \frac{1}{(1+t^2)(1+\mu_2t^2)}\right),
\ee
which expresses the invariant sector as the coefficient before the bracket, and the contributions of the trivial and non trivial expressions to the first and second part in the bracket, respectively.

Inspecting \eref{HWGf2k} after setting $q=1$ since the $U(1)$ is not an isometry for the orbifold, and comparing to \eref{branch86} we find that the terms $t^2$ and $\mu_k t^k$ are non trivial under $\BZ_2$ while the other terms are invariant. Again we can form 3 invariants under the $\BZ_2$ action consisting of $t^4$, $\mu_k t^{k+2}$, and $\mu_k^2 t^{2k}$, with a relation $\mu_k^2 t^{2k+4}$. Putting all the terms together we find
\be\label{HWGI2k}
HWG(\CH_\infty)_{n=2, k} = \PE \left[\sum_{i=1}^{k-1} \mu_i \mu_{2k-i} t^{2i} + \mu_k t^k + t^4  + \mu_k t^{k+2} + \mu_k^2 t^{2k} - \mu_k^2 t^{2k+4} \right],
\ee
where $\mu_i$ are highest weight fugacities for $SU(2k)$.
\eref{HWGI2k} coincides with \eref{HWG226} when $k=2$.
This is a result for any value of $k$ and it has been tested using explicit computations from the quiver in \eref{mirrorUkinfinite} for $n=2$ with full agreement. We conclude that
\be
\CH_\infty \left( {\sf Q}_{2,k} \right) = \CH_f \left( {\sf Q}_{2,k} \right) / S_2 .
\ee
This constitutes another nontrivial test of Conjectures \eref{discrete} and \eref{Ohmori}.

It is important to test the conjectures on non Abelian discrete groups $S_n$ and we turn to this next.

\subsection{$n$ M5-branes on a $\BC^2/\BZ_2$ singularity}
In the two examples above the number of M5 branes is $n=2$ and correspondingly there are two phases: separated and coincident, or finite and infinite gauge coupling, respectively. In the next set of examples we have as many phases as partitions of $n$, hence each phase is going to be labeled by a partition. We therefore change notation from subscripts $f$ and $\infty$ to subscripts given by partitions. The phase where all gauge couplings are finite is denoted by $\{1^n\}$ and the phase where all gauge couplings are infinite is denoted by $\{n\}$. The quiver \eref{mirrorSUk} for $k=2$ at the separated phase takes the form
\be \label{quiverk21n}
{\sf F}_{\{1^n\},2} =~~\node{}{1}-\node{\overbrace{{\overset{1~{\node{}{}}\quad \ldots\quad{\node{}{}}~1}{\setminus\cdots /}}}^{n~\text{nodes}}}{2}-\node{}{1}~,
\ee
with Higgs branch dimension $n+3$ and a global symmetry of $SU(2)^{n+2}$. There is an enhancement of the discrete global symmetry from $S_n$ to $S_{n+2}$, and discrete gauging of subgroups for this special case can produce interesting moduli spaces, with accidental enhanced global symmetry, as discussed in the next two subsections. The HWG for any $n$ was computed in \cite{Hanany:2010qu} and takes the form
\be
HWG\left(\CH_{\{1^n\}}\right)_{n, k=2} = \PE \left[t^2 \sum_{i=1}^{n+2} \nu_i^2 + t^4  + \left(t^n+t^{n+2}\right)\prod_{i=1}^{n+2}\nu_i - t^{2n+4} \prod_{i=1}^{n+2}\nu_i^2 \right],
\ee
where $\nu_i$ is the fugacity for the highest weights of the $i$-th $SU(2)$. The action of $S_{n+2}$ is evident on this HWG and discrete quotients naturally reduce the number of global $SU(2)$ symmetries as they transform in the fundamental representation of $S_{n+2}$. There is one diagonal $SU(2)$ which transforms in the trivial representation, for each part of the partition. Hence such a global symmetry survives the projection by the subgroup of $S_{n+2}$. As a result, the number of surviving $SU(2)$ groups in each phase is the number of parts of the partition. This is a derivation of the observation made above that the global symmetry when all inverse gauge couplings are 0 is $SU(2)^3$, as it corresponds to the partition $\{n,1^2\}$ of $n+2$.
Next we turn to special cases.
\subsection{$SO(8)$ to $G_2$}
\label{subsectionG2}
For this case we again set $n=2$ and pick the partition $\{3,1\}$ of 4. The quiver takes the form
\be \label{31quiver}
{\sf F}_{\{3,1\}} = ~ ~ \node{}{1}-\node{\overset{\cap}{\ver{}{\,\,3}}}{2}~, 
\ee
and the global symmetry was identified to be $G_2$ in \cite{Gaiotto:2012uq} and the HWG was computed in Equation 3.37 of \cite{Cremonesi:2014vla} to be
\be
HWG\left(\CH_{\{3,1\}}\right)_{2, 2} = \PE \left[\mu_2 t^2 + \mu_1^2 t^4 + \mu_1^3 t^6 + \mu_2^2 t^8 + \mu_1^3\mu_2 t^{10} - \mu_1^6 \mu_2^2 t^{20} \right],
\ee
where $\mu_1, \mu_2$ are fugacities for highest weights of $G_2$. The corresponding moduli space is identified \cite{Hanany:2017ooe} to be the closure of the sub regular orbit of $G_2$,
\be
\CH_{\{31\}} = \overline{{\rm s.reg}_{G_2}}
\ee
By recalling that the moduli space for finite coupling is the closure of the minimal orbit of $SO(8)$
\be
\CH_{\{1^4\}} = \overline{{\rm min}_{SO(8)}}
\ee
and the classic result by Kostant and Brylinski \cite{1992math......4227B}
\be
\overline{{\rm s.reg}_{G_2}} = \overline{{\rm min}_{SO(8)}} / S_3,
\ee
We recover the important result
\be \label{CBrel}
\CH_{\{31\}} = \CH_{\{1^4\}} / S_3 .
\ee

The Coulomb branch of the quiver \eref{31quiver} can be related to the Higgs branch of a 6d SCFT closely related to the one on $3$ M5-branes on a $\BC^2/\BZ_2$ singularity. Particularly, the latter SCFT can be described by a Type IIA brane configuration with $2$ D6-branes intersected by $3$ NS5-branes, as reviewed in the beginning of this section. 

Consider the two semi-infinite D6-branes on one side. These can be made to end on D8-branes. When allowing them to end on the D8-branes we have a choice on whether they end on separate D8-branes or on the same D8-brane. The 6d SCFT associated with $3$ M5-branes on a $\BC^2/\BZ_2$ singularity is related to the former choice while the latter describes a different SCFT. 

This new SCFT can be reached from the one on the $3$ M5-branes on a $\BC^2/\BZ_2$ singularity by going on the Higgs branch of that SCFT, the one described in the brane configuration by performing a KP transition \cite{Cabrera:2016vvv} breaking off to infinity the D6-brane part between the two D8-branes.

As we argue in Section \eref{relationship}, Conjecture \eref{Ohmori} can be generalized to also include these types of 6d SCFTs, and when applied to this case leads to the claim that the Coulomb branch of quiver \eref{31quiver} is the Higgs branch of this SCFT. Now, the 6d SCFT on $3$ M5-branes on a $\BC^2/\BZ_2$ singularity has a low-energy gauge theory description as an $SU(2)\times SU(2)$ gauge theory with a bifundamental hyper and two fundamental hypers for each of the two $SU(2)$ gauge groups, given in quiver \eref{linearquiver} for $n=3, k=2$. 

The Higgs branch motion leading to the new SCFT from this one, is described in the low-energy gauge theory by giving a vev to the meson of one of the $SU(2)$ gauge groups. This causes a flow under which the low-energy gauge theory flows to an $SU(2)$ gauge theory with four fundamental hyper multiplets and two tensor multiplets, as while the vector multiplet associated with the other $SU(2)$ gauge group is Higgsed down, the tensor multiplet associated with it remains.    

We now see that this actually provides a confirmation of Conjecture \eref{discrete}, in a very special way! The 6d SCFT whose Higgs branch is associated with the Coulomb branch of quiver \eref{31quiver} is given by $3$ coincident M5-branes. However, when these are separated we expect the Higgs branch to be described by just one of the $SU(2)$ gauge groups with four fundamental hyper multiplets. Relation \eref{CBrel} is exactly in accordance with Conjecture \eref{discrete}.

Finally, we wish to remark that this is actually a special case of Conjecture \eref{Anton}, and will be elaborated upon in \cite{HZ}.

\subsection{$SO(8)$ to $SU(3)$}
\label{subsectionSU3}
For this case we set $n=2$ and pick the partition $\{4\}$ of 4. The quiver takes the form
\be \label{4quiver}
{\sf F}_{\{4\}} = ~ ~ \node{\overset{\cap}{\ver{}{\,\,4}}}{2}~, 
\ee
The global symmetry is $SU(3)$ \cite{HZ}, and the moduli space is confirmed to be an $S_4$ orbifold of the finite coupling moduli space \cite{HZ},
\be \label{CBrel1}
\CH_{\{4\}} = \CH_{\{1^4\}} / S_4
\ee

Similarly to the previous case the Coulomb branch of \eref{4quiver} can also be related to the Higgs branch of a 6d SCFT, now closely related to the one on $4$ M5-branes on a $\BC^2/\BZ_2$ singularity. The idea is similar to the previous case just that now we take the Higgs branch flow on both pairs of semi-infinite D6-branes. 

In the low-energy gauge theory, we now have an $SU(2)\times SU(2) \times SU(2)$ gauge theory with bifundamental hypers and two fundamental hypers for the $SU(2)$ gauge groups at the ends of the quiver, as in \eref{linearquiver} for $n=4, k=2$. The Higgs branch limit is then described by giving a vev to an $SU(2)$ meson, now at both ends of the quiver. This should cause the low-energy theory to flow again to an $SU(2)$ gauge theory with four fundamental hyper multiplets, but now with three tensor multiplets.

  By the generalization of Conjecture \eref{Ohmori}, that we argue in Section \eref{relationship}, the Coulomb branch of quiver \eref{4quiver} should be equal to the Higgs branch of this 6d theory when the $4$ M5-branes coincide. However, when these are separated we expect the Higgs branch to be described just by one of the $SU(2)$ gauge groups (middle one) with four fundamental hyper multiplets. The relation \eref{CBrel1} is then exactly in accordance with Conjecture \eref{discrete}.

\section{Small instantons, A type singularity, and coincident M5 branes}
\label{Smallinstantons}
In the next class of examples we consider a mix of A type (discrete gauging) and D type (small instanton KP transition) behavior. As our example we use 6d SCFTs that can be constructed via a brane configuration consisting of M5 branes in the presence of an M9 plane and a $\BC^2/\BZ_m$ singularity. Naturally, there are many cases one can consider, and we expect the structure exhibited here to apply for all of them. However, here we shall concentrate on two families, which are relatively simple yet exhibit interesting structure.

As our first case we consider the system of $n$ M5 branes in the presence of an M9 plane and a $\BC^2/\BZ_2$ singularity. There are 3 distinct SFTs associated with this configuration differing by  the action (or embedding) of the $\BZ_2$ on the $E_8$ symmetry \cite{Heckman:2015bfa}. Here we concentrate on the case where the $\BZ_2$ action on $E_8$ preserves its $SO(16)$ maximal subgroup. The global symmetry of the SCFT is $SU(2)\times SU(2)\times SO(16)$, where one $SU(2)$ comes from the $\BZ_2$ singularity (A $\BZ_m$ orbifold in M theory leads to an $SU(m)$ global symmetry on the M5 brane), another from the isometry of $\BC^2/\BZ_2$ and the $SO(16)$ from the commutant of $\BZ_2$ in $E_8$. The gauge theory on the tensor branch, where all gauge couplings are finite, can be derived from a Type I$'$ background with $n$ NS5 branes and 2 D6 branes next to an O8$^-$ plane with 8 D8 branes \cite{Hanany:1999sj}.  The quiver theory reads
\be \label{linearquiverM9}
{\sf M}_{n,2} = ~~\underbrace{\node{\redsqver{}{\,\,SO(4)}}{2} - \node{}{2} - \cdots - \node{}{2} - \node{}{2}}_{n - 1~\text{\comment{$(k)$} nodes}}  - \bnode{\redsqver{}{~SO(16)}}{Sp(1)} ~,
\ee
where each gauge node is $SU(2)$ and the right node is of $Sp$ type due to the presence of the O8$^-$ plane. The Higgs branch when all gauge couplings are finite has dimension $n+16$ and a global symmetry $SU(2)^{n+1}\times SO(16)$, where we note that there is an $SU(2)$ global symmetry factor per each edge between 2 gauge nodes, as the vector representation of $SO(4)$ is real. While this global symmetry is evident from the quiver, it is significantly greater than that of the SCFT. This is similar to the case of M5 branes on a $\BC^2/\BZ_2$ singularity which is discussed in the previous section, and can again be understood through the phenomenon of discrete gauging.  

Less evident is an $S_n$ discrete symmetry which is visible in the 3d Coulomb branch quiver for the finite gauge couplings phase. Indeed, this theory has a multitude of phases which can best be described by a partition of $n$. Define $\{n_i\}_{i=0}^{l}$ to be a partition of $n = \sum_{i=0}^{l} n_i$, where $l+1$ is the number of parts in the partition of $n$. We set $n_0$ to be the number of M5 branes inside the M9 plane, and the other $n_i$ to count coincident M5 branes outside of the M9 plane, located in distinct positions $x_i$ along the M theory interval. $n_0$ is allowed to be 0, while $n_i$ are non negative integers. Set $p_n$ to be the number of partitions of $n$, Then the total number of phases is
\be
\# {\rm phases} = \sum_{i=0}^n p_i .
\ee
Each of these phases has a different Higgs branch which is best described by a Coulomb branch of a 3d $\CN=4$ quiver. In the following we give the most general such quiver and point out some special cases. By convention a circle represents a balanced (twice rank is sum of neighbors) node, while a grey node is unbalanced.
\begin{itemize}
\item
When all M5 branes are inside the M9 plane, $n=n_0$, the corresponding 3d quiver looks like \cite{Mekareeya:2017jgc}
\be \label{SO4SO16Qinfinite}
\node{}{1} - ~\gnode{}{2} - \node{}{n+2}- \node{}{2n+2}- \node{}{3n+2}- \node{}{4n+2}- \node{}{5n+2} -\node{\ver{}{3n+1}}{6n+2}-\node{}{4n+1}-\gnode{}{2n}~, 
\ee
with Higgs branch dimension $30n+16$, and isometry $SU(2)\times SU(2) \times SO(16)$, as expected of the 6d SCFT. Here we note that while there are 2 unbalanced nodes, which typically give rise to a $U(1)$ isometry, there is a further enhancement to $SU(2)$ due to the special structure of the quiver.
% additional symmetry is  %The Higgs branch dimension grows by $29n$ suitable to $n$ small instanton transitions.
\item
When all M5 branes are away from the M9 plane and are separated, $n_0=0, n_i=1$, the 3d quiver takes the form
\be \label{SO4SO16Qtrivial}
\node{}{1} ~ - \gnode{\overbrace{{\overset{1~\overset{}{\node{}{}}\quad \cdots\quad\overset{}{\node{}{}}~1}{\setminus\cdots /}}}^{n~\text{nodes}}}{2} - \node{}{2}- \node{}{2}- \node{}{2}- \node{}{2}- \node{}{2} -\node{\ver{}{1}}{2}-\node{}{1}~,
\ee
where the grey (unbalanced) node has a bouquet of $n$ $U(1)$ nodes attached to it. All other nodes are balanced. The Higgs branch dimension is $n+16$. Indeed, the difference in dimensions is equal $29n$ suitable for $n$ small instanton transitions ($n$ affine $E_8$ quiver subtractions \cite{Cabrera:2018ann, Hanany:2018uhm}). The global symmetry is $SU(2)^{n+1}\times SO(16)$, as is evident from the Dynkin diagram formed by the balanced nodes. In addition there is a discrete symmetry of $S_{n+1}\times S_2$. For this case we can compute the HWG of the Coulomb branch with the help of Equation (7.1) of \cite{Hanany:2010qu}

\be
HWG = \PE \left[t^2 \sum_{i=1}^{n+1} \nu_i^2 + \mu_2 t^2 + (1+\mu_1^2) t^4  + \mu_1 (t^{n+1}+t^{n+3}) \prod_{i=1}^{n+1}\nu_i - \mu_1^2 t^{2n+6}\prod_{i=1}^{n+1}\nu_i^2 \right],
\ee
where $\nu_i$ are fugacities for the highest weights of the $SU(2)$ symmetries and $\mu_i$ are fugacities for the highest weights of the $SO(16)$ global symmetry.
\item
The most general configuration where the $n_i$ are general has a 3d quiver which takes the form
\be \label{SO4SO16Qgeneral}
\node{}{1} - \gnode{\overbrace{{\overset{n_1~\overset{\cap}{\node{}{}}  \quad\cdots\quad\overset{\cap}{\node{}{}}~n_{l}}{\setminus\cdots /}}}^{l~\text{nodes}}}{2} - \node{}{n_0+2}- \node{}{2n_0+2}- \node{}{3n_0+2}- \node{}{4n_0+2}- \node{}{5n_0+2} -\node{\ver{}{3n_0+1}}{6n_0+2}-\node{}{4n_0+1}-\gnode{}{2n_0}~, 
\ee
with a bouquet of adjoint $n_i$ nodes attached to the grey node. The global symmetry is $SU(2)^{l+1}\times SO(16)$. The dimension of the Higgs branch is $29n_0+n+16$ which demonstrates that it depends on the number of M5 branes inside the M9 plane and the total number of M5 branes, but not on the coincidence.
\end{itemize}
As in the previous cases, we defer arguments in favor of these quivers to Section \eref{relationship}.

This family of quivers has inter relations between the quivers which involve KP transitions of minimal $E_8$ type as in \cite{Hanany:2018uhm} and discrete gauging as in this paper. The KP transition is along the lines of quiver subtractions as discussed in \cite{Cabrera:2018ann}.
In detail, these relations are as follows:
\paragraph{Small instanton and the KP transition}
When a single M5 branes leaves the M9 plane the partition of $n$ changes such that the number of parts grows by 1, and $n_0$ is reduced by one. Denote the first partition by $\lambda = \{n_i\}$ and the new one by $\lambda' = \{n_0-1;n_i, 1\}$. Denote the corresponding 3d quivers by ${\sf M}_\lambda$ and ${\sf M}_{\lambda'}$.
Using the results of \cite{Cabrera:2018ann, Hanany:2018uhm} we have the relation between Coulomb branches, which naturally induces a relation between the 6d Higgs branches,
\be
\CC\left({\sf M}_\lambda\right) \supset \CC\left({\sf M}_{\lambda'}\right) ,
\ee
where the transverse slice to $\CC\left({\sf M}_{\lambda'}\right)$ inside $\CC\left({\sf M}_\lambda\right)$ is $\overline{{\rm min}_{E_8}}$, the minimal nilpotent orbit of $E_8$.
\paragraph{Discrete gauging}
When $n_i$ M5 branes coincide away from the M9 plane we have a discrete gauging of $S_{n_i}$. Denote the almost trivial partition where $n_0$ M5 branes are inside the M9 plane and the remaining $n-n_0$ M5 branes are separated away from the M9 plane, $n_i=1, i\not=0$, by $\lambda_0$. Then by Conjecture \eref{Anton} we have the relation
\be
\CC\left({\sf M}_\lambda\right) = \CC\left({\sf M}_{\lambda_0}\right)/\prod_{i=1}^l S_{n_i} ,
\ee
which naturally induces a relation between these 6d Higgs branches.

\subsection{$E_8 \supset SU(9)$ sequence -- Quiver Gymnastics}
We now turn to discuss the following one parameter family of 3d $\CN=4$ quivers, which was studied in \cite{Zafrir:2015rga, Hayashi:2015zka}, however we take a new look using the ideas of quiver subtractions \cite{Cabrera:2018ann}, the KP small instanton transition \cite{Hanany:2018uhm}, and discrete gauging as in this paper. 

As the name suggests, the 6d theories of interest here correspond to cases where the $Z_m$ orbifold action on $E_8$ preserves its $SU(9)$ subgroup. For this to be possible we must have $m=3(k-2)$, where $k$ is an integer \cite{Heckman:2015bfa}. Here we shall concentrate on the minimal number of M5-branes possible for this case, which is $n=\frac{k+1}{3}$ rounded down. For this case the $SU(9)$ global symmetry from $E_8$ merges with the $SU(3k-6)$ expected from the orbifold, leading to a 6d SCFT with $SU(3k+3)$ global symmetry. 

\begin{figure}
\center
\includegraphics[width=0.8\textwidth]{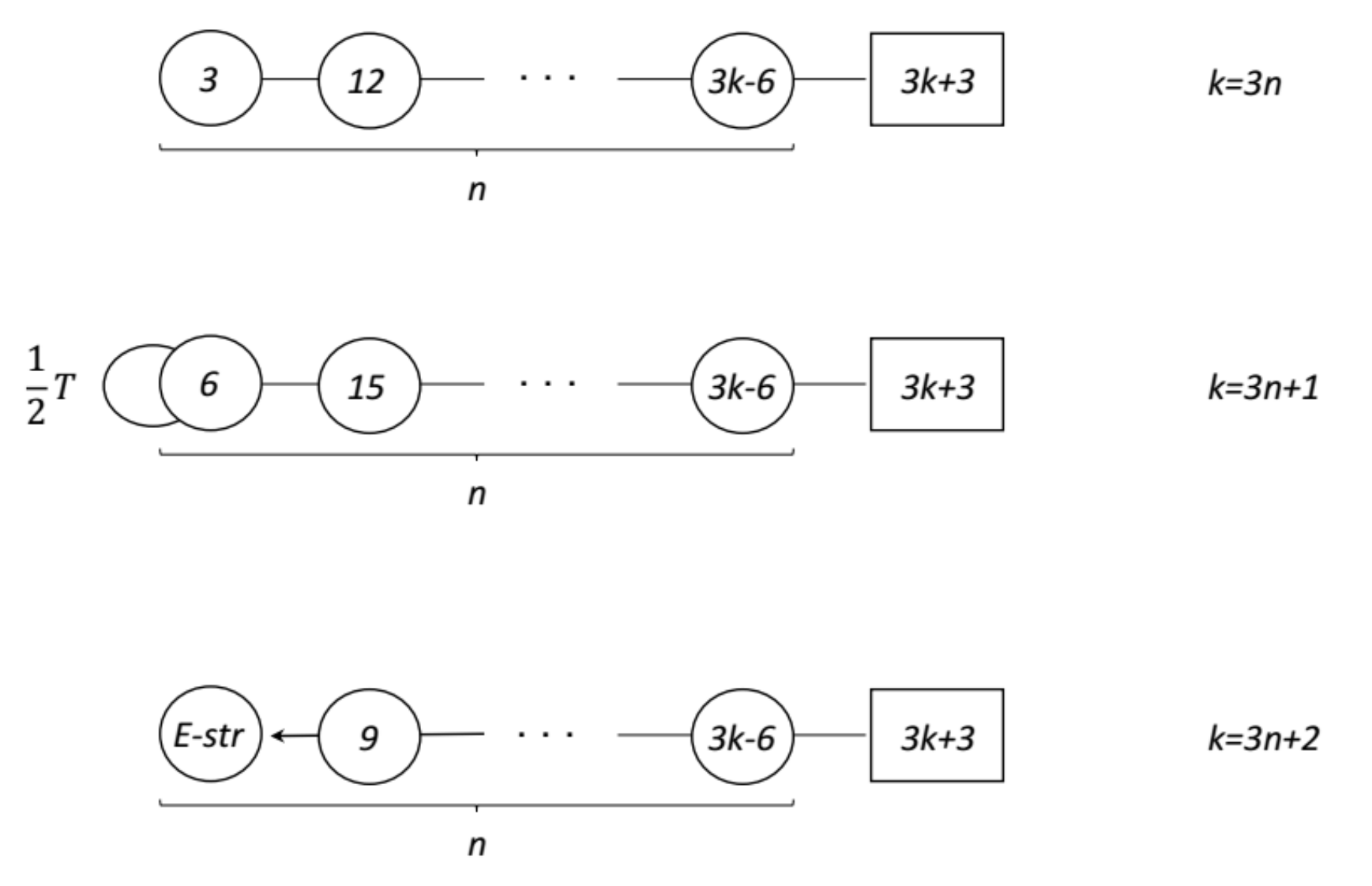} 
\caption{The 6d quivers of the low-energy theory on the tensor branch, associated with the family of 6d SCFTs studied in this subsection. The form of the quiver differs depending on the value of $k$ mod $3$. All groups are of type $SU$ and in the second quiver the line from $SU(6)$ to itself stands for a half-hyper in the three index antisymmetric representation. Also the leftmost circle in the bottom quiver stands for the rank $1$ E-string theory, which is connected to the neighboring $SU(9)$ gauge group via gauging of $SU(9) \subset E_8$.}
\label{6dQuivers}
\end{figure}

From the results of \cite{Mekareeya:2017jgc}, the 6d quivers on the tensor branch can be constructed. These are shown in Figure \ref{6dQuivers}. As we see below, these 6d quivers appear in a rather non-trivial way in the study of the Higgs branches of the 6d SFTs through the associated 3d quivers, which we next turn to study. We again defer arguments in favor of these quivers to Section \eref{relationship}.      

The 3d quivers in this case take the form,
\be \label{E8SU9}
{\sf Q}_{k} ~~ = \node{}{1} - \node{}{2} -\cdots - \node{}{3k-7} - \node{}{3k-6} - \node{}{3k-5} - \node{}{3k-4}- \node{}{3k-3}- \node{}{3k-2}- \node{}{3k-1} -\node{\gver{}{k+1}}{3k}-\node{}{2k}-\node{}{k}~, 
\ee
where the right nodes are explicitly spelled out for an easy subtraction of the affine $E_8$ quiver. This family of quivers belongs to the class of minimally unbalanced quivers \cite{CHZ}, where the only unbalanced node is depicted in gray with an imbalance of $k-2$. The dimension of the Coulomb branch is $k(9k+11)/2$ and the global symmetry is $SU(3k+3)$. The imbalance indicates that cases $k=1$ and $k=2$ are special, which we now review.
For $k=1$ this node has an imbalance of $-1$ indicating that there is a free sector. The node is attached to the 3-rd node indicating that the corresponding monopole operators transform in the [00100] representation of $SU(6)$ of dimension 20. We conclude that there are 20 such monopole operators of spin $1/2$ under $SU(2)_R$, generating a copy of $\BH^{10}$ but since the dimension is 10, the whole Coulomb branch is free,
\be
\CC\left( \node{}{1} - \node{}{2} -\node{\gver{}{2}}{3}-\node{}{2}-\node{}{1}~ 
\right) = \BH^{10}~.
\ee
As the global symmetry of this space is $Sp(10)$ we get an interesting embedding of $SU(6)\subset Sp(10)$ under which the fundamental of $Sp(10)$ becomes the pseudo real representation $[00100]$ of $SU(6)$.

For $k=2$ the grey node is balanced and the global symmetry gets enhanced from $SU(9)$ to $E_8$. Indeed the simplest monopole operators coming from this node transform in the $(\wedge^3+\wedge^6)[10000000]$ of $SU(9)$, each with dimension 84, and together with the adjoint of $SU(9)$ of dimension $80$ form the adjoint representation of $E_8$. As the dimension of the Coulomb branch is 29 we identify the singularity as the closure of the minimal nilpotent orbit of $E_8$. Indeed the quiver is nothing but the affine $E_8$ Dynkin diagram,
\be
\CC\left( \node{}{1} - \node{}{2} - \node{}{3} - \node{}{4} - \node{}{5} -\node{\ver{}{3}}{6}-\node{}{4}-\node{}{2}~ 
\right) = \overline {{\rm min}_{E_8}}~.
\ee

Next we study the case $k=3$. Here we notice that the $k=2$ quiver is a sub quiver, as is the case for any $k>2$, hence perform a quiver subtraction \cite{Cabrera:2018ann} to get a small instanton transition. It is crucial to realize that not every quiver is allowed for subtraction due to the 6d gravitational anomaly conditions, which allow a trade of 1 tensor multiplet by 29 hypermultiplets. Therefore we need to be careful in picking the subtracted quiver. As we are subtracting a quiver which has a 29 dimensional Coulomb branch, it is a sufficient condition for the resulting quiver to have a Coulomb branch which is a Higgs branch of some 6d theory,
\be
\label{312fin}
{\sf Q}_3 - {\sf Q}_2 = \node{}{1} - ~\node{}{2} - \node{\gver{1}{}}{3} - \node{}{3} - \node{}{3}- \node{}{3}- \node{}{3}- \node{}{3} -\node{\gver{}{1}}{3}-\node{}{2}-\node{}{1}~
\ee
This is a familiar quiver \cite{Hanany:1996ie}, satisfying the relation
\be
\CH_f\left( ~ \node{\wver{}{12}}{3} \quad\right) = \CC\left ( {\sf Q}_3 - {\sf Q}_2 \right) ,
\ee
from which we conclude from the notion of transverse slice explained in \cite{Cabrera:2018ann} that
\be
\label{312inf}
\CH_\infty\left( ~ \node{\wver{}{12}}{3} \quad\right) = \CC\left({\sf Q}_3\right).
\ee
Also note that this theory is free from gauge anomalies in 6d \cite{Danielsson:1997kt}. \eref{312inf} is an important observation which says that the Higgs branch of the 6d $\CN=(1,0)$ supersymmetric $SU(3)$ gauge theory with 12 flavors undergoes a small instanton transition at infinite coupling and the new Higgs branch is given by the Coulomb branch of the 3d quiver ${\sf Q}_3$, reproducing a prediction from \cite{Zafrir:2015rga} which is using a different technique. The theory has two phases: the finite coupling phase with Higgs branch given by the Coulomb branch of quiver \eref{312fin}, and the infinite coupling phase with the quiver ${\sf Q}_3$. The number of phases of this theory is $p_0+p_1=2$, where $p_n$ is the number of partitions of $n$. We see below how this count of phases is generalized to higher  values of $k$.

The next case is $k=4$. We expect it to be a phase of some theory, and in order to compute which theory, we use a quiver subtraction. Again we are restricted by subtraction of quivers which satisfy anomaly cancellation. The difference takes the form
\be
\label{615fin}
{\sf Q}_4 - {\sf Q}_2 = \node{}{1} - \node{}{2} - \node{}{3} - \node{}{4} - \node{}{5} - \node{\gver{1}{}}{6} - \node{}{6} - \node{}{6}- \node{}{6}- \node{}{6}- \node{}{6} -\node{\gver{}{2}}{6}-\node{}{4}-\node{}{2}~
\ee
This is not a familiar quiver, so we need a further subtraction to get something familiar. We demonstrate below that this subtraction gives a new mirror prediction. First note that 7 nodes have 6, hence we expect the mirror to contain a 6. We recall ${\sf Q}_1$ which represents 20 free half hypers in the [00100] of $SU(6)$. In turn we recall from \cite{Danielsson:1997kt} that $SU(6)$ is the only gauge group which couples to such a representation, free of gauge anomalies, and the number of flavors is uniquely fixed to be 15. We proceed by subtracting ${\sf Q}_1$ of \eref{E8SU9} to get
\be
\label{615finno20}
{\sf Q}_4 - {\sf Q}_2 - {\sf Q}_1 = \node{}{1} - \node{}{2} - \node{}{3} - \node{}{4} - \node{}{5} - \node{\gver{1}{}}{6} - \node{}{6} - \node{}{6} - \node{}{6} - \node{\gver{1}{}}{6} - \node{}{5} - \node{}{4} - \node{}{3} - \node{}{2}-\node{}{1}~
\ee
This quiver is certainly of familiar form \cite{Hanany:1996ie} giving the 3d mirror to be $SU(6)$ with 15 flavors, however, from the table of \cite{Danielsson:1997kt} we see that such a theory is anomalous in 6d unless one adds 20 half hypers in the [00100] of $SU(6)$, which is precisely what we have for \eref{615fin}. We are therefore led to the following 3d mirror conjecture:
\begin{conj}
The 3d mirror of $SU(6)$ gauge theory with 15 flavors and 1/2 hyper in the [00100] of dimension 20 is given by quiver \eref{615fin}.
\end{conj}
So much for the 3d mirror, but there is a crucial result to obtain in 6d. The same theory has two phases, $p_0+p_1 = 2$, given by finite coupling and by infinite coupling. The Higgs branch at finite coupling is given by the Coulomb branch of quiver \eref{615fin},
\be
\CH_f\left( ~ \node{\wver{}{15}}{6 ~{\rm with} ~ \frac{1}{2}\wedge^3} \quad\right) = \CC\left ( {\sf Q}_4 - {\sf Q}_2 \right),
\ee
of dimension 65, and the Higgs branch at infinite coupling is given by the Coulomb branch of the ${\sf Q}_4$ quiver.
\be
\CH_\infty\left( ~ \node{\wver{}{15}}{6 ~{\rm with} ~ \frac{1}{2}\wedge^3} \quad\right) = \CC\left ( {\sf Q}_4  \right) ,
\ee
of dimension 94. As above, at infinite coupling the theory undergoes a small instanton transition and the Higgs branch grows by 29 new (quaternionic) flat directions.

We proceed with $k=5$. The new feature is that one can perform two affine $E_8$ subtractions. Correspondingly, the number of phases of the 6d theory are $p_0+p_1+p_2=4$. Let us collect the two quivers resulting from subtraction,
\be
\label{91811}
{\sf Q}_5 - {\sf Q}_2 = \node{}{1} - \node{}{2} -~\cdots~- \node{}{8} - \node{\gver{1}{}}{9} - \node{}{9} - \node{}{9}- \node{}{9}- \node{}{9}- \node{}{9} -\node{\gver{}{3}}{9}-\node{}{6}-\node{}{3}~
\ee
and
\be
\label{918011}
{\sf Q}_5 - {\sf Q}_2 - {\sf Q}_2 = \node{}{1} - \node{}{2} -~\cdots~- \node{}{8} -\node{{\overset{1~{\gnode{}{}}~~{\gnode{}{}}~1}{\setminus /}}}{9} - \node{}{8}- ~ \cdots ~ - \node{}{2}-\node{}{1}~
\ee
At this point we find, using the 3d mirror which was computed in \cite{Hanany:1996ie}, the anomaly free theory of $SU(9)$ with 18 flavors with a gauge coupling,  coupled to an additional tensor multiplet (its presence is denoted by an additional node with label 0). Correspondingly, this theory has 4 different phases, each with a different Higgs branch. The pattern of the different phases follows the pattern of Levy subgroups of $B, C,$ or $D$ type algebras, as in \cite{Hesselink1978}, or expressed in physical terms, the pattern of the adjoint Higgs mechanism in D$p$ branes next to $Op$ planes. We therefore denote the different phases by partition data of the form $\{n_0; n_i\}_{i=1}^l$, where $n_0$ is the number of small instanton transitions from the phase where all BPS strings have finite tension (by analogy to $n_0$ D$p$ branes on the $Op$ plane), and $n_i$ for $i=1\ldots l$ denote tensionless strings away from the origin of the tensor branch (by analogy to $n_i$ coincident D$p$ branes away from the $Op$ plane). The 4 phases are as follows.
\begin{enumerate}
\item The phase $\{0;1,1\}$ where all BPS string tensions are non zero. The Higgs branch is given by the Coulomb branch of \eref{918011}
\be
\label{Higgs918011}
\CH_{\{0;1,1\}}\left( ~ \node{}{0} - \node{\wver{}{18}}{9} \quad\right) = \CC \eref{918011},
\ee
of dimension 82.
\item The phase $\{1;1\}$ with one scalar in the tensor multiplet at the origin, generating a small instanton transition, and one scalar away from the origin. The Higgs branch is given by the quiver \eref{91811},
\be
\CH_{\{1;1\}}\left( ~ \node{}{0} - \node{\wver{}{18}}{9} \quad\right) = \CC \eref{91811},
\ee
of dimension 111.
\item The phase $\{2;0\}$ with 2 scalars at the origin. The Higgs branch is given by the Coulomb branch of ${\sf Q}_5$ in \eref{E8SU9}
\be
\CH_{\{2;0\}}\left( ~ \node{}{0} - \node{\wver{}{18}}{9} \quad\right) = \CC \left({\sf Q}_5\right),
\ee
of dimension 140.
\item The phase $\{0; 2\}$ with 2 scalars away from the origin but with equal value, leading to a tensionless string. The Higgs branch is an $S_2$ quotient of the Higgs branch of \eref{Higgs918011} and is given by the quiver
\be
\label{91802}
\CH_{\{2;0\}}\left( ~ \node{}{0} - \node{\wver{}{18}}{9} \quad\right) = \left( \node{}{1} - \node{}{2} -~\cdots~- \node{}{8} - \node{\overset{\cap}{\ver{}{\,\,2}}}{9} - \node{}{8}- ~ \cdots ~ - \node{}{2}-\node{}{1}~ \right) .
\ee
\end{enumerate}

Next we turn to the general $k$ case. We notice that there are 3 subfamilies given by the value of $k ~ \mod ~ 3$. We treat each sub family separately, but before this, let us perform $j$ small instanton transitions, with the relation $3j+3n_0=k+1$, and write the most general quiver ${\sf S}_{k,j,\{n_i\}}$ for $n_i$ coincident branes such that $\sum_{i=1}^l n_i = j$.
\bea \label{E8SU9any}
&{\sf S}_{k,j,\{n_i\}} = & \\ \nonumber
&\node{}{1} - ~\node{}{2} -\cdots - \node{}{3k-7} - & \gnode{\overbrace{{\overset{n_1~\overset{\cap}{\node{}{}}  \quad\cdots\quad\overset{\cap}{\node{}{}}~n_{l}}{\setminus\cdots /}}}^{l~\text{nodes}}}{3k-6} - \node{}{3k-j-5} - \node{}{3k-2j-4}- \cdots
%\node{}{3k-3j-3} - \node{}{3k-4j-2}
- \node{}{3k-5j-1} -\node{\ver{}{k-3j+1}}{3k-6j}-\node{}{2k-4j}-\node{}{k-2j}~, 
\eea
which satisfies orbifold relations between the moduli spaces,
\be
\CC\left({\sf S}_{k,j,\left\{{n_i}\right\}}\right) = \CC\left({\sf S}_{k,j,\left\{{1^j}\right\}}\right) / \prod_i S_{n_i}
\ee

The sub family with $k=0 ~ \mod ~ 3$ has a phase $\{0; 1^{\frac{k}{3}}\}$ where all BPS strings have finite tension, which sets the low energy quiver and the relation
\be
\CH_{\left\{0; 1^{\frac{k}{3}}\right\}}\left( ~ \node{}{3} - \node{}{12} - ~ \cdots ~ - \node{}{3k-15} - \node{\wver{}{3k+3}}{3k-6} \qquad\right) = \CC\left({\sf S}_{k,\frac{k}{3},\left\{1^{\frac{k}{3}}\right\}}\right).
\ee
The number of phases of the theory is given by $\sum_{i=0}^{\frac{k}{3}}p_i$ and each phase has a Higgs branch given by
\be
\CH_{\left\{n_0; n_i\right\}}\left( ~ \node{}{3} - \node{}{12} - ~ \cdots ~ - \node{}{3k-15} - \node{\wver{}{3k+3}}{3k-6} \qquad\right) = \CC\left({\sf S}_{k,j,\left\{{n_i}\right\}}\right) %= \CC\left({\sf S}_{k,j,\left\{{1^j}\right\}}\right) / \prod_i S_{n_i}
.
\ee
The sub family with $k=1 ~ \mod ~ 3$ has a phase $\{0; 1^{\frac{k-1}{3}}\}$ where all BPS strings have finite tension, which sets the low energy quiver and the relation
\be
\CH_{\left\{0; 1^{\frac{k-1}{3}}\right\}}\left( \qquad \node{\wver{\frac{1}{2}\wedge^3~}{}}{6} - \node{}{15} - ~ \cdots ~ - \node{}{3k-15} - \node{\wver{}{~3k+3}}{3k-6} \qquad\right) = \CC\left({\sf S}_{k,\frac{k-1}{3},\left\{1^{\frac{k-1}{3}}\right\}}\right).
\ee
The number of phases of the theory is given by $\sum_{i=0}^{\frac{k-1}{3}}p_i$ and each phase has a Higgs branch given by
\be
\CH_{\left\{n_0; n_i\right\}}\left( \qquad \node{\wver{\frac{1}{2}\wedge^3~}{}}{6} - \node{}{15} - ~ \cdots ~ - \node{}{3k-15} - \node{\wver{}{~3k+3}}{3k-6} \qquad\right) = \CC\left({\sf S}_{k,j,\left\{n_i\right\}}\right) = \CC\left({\sf S}_{k,j,\left\{{1^j}\right\}}\right) / \prod_i S_{n_i} .
\ee
The sub family with $k=2 ~ \mod ~ 3$ has a phase $\{0; 1^{\frac{k+1}{3}}\}$ where all BPS strings have finite tension, which sets the low energy quiver and the relation
\be
\label{k2finite}
\CH_{\left\{0; 1^{\frac{k+1}{3}}\right\}}\left( ~ \node{}{0} - \node{}{9} - ~ \cdots ~ - \node{}{3k-15} - \node{\wver{}{3k+3}}{3k-6} \qquad\right) = \CC\left({\sf S}_{k,\frac{k+1}{3},\left\{1^{\frac{k+1}{3}}\right\}}\right).
\ee
The number of phases of the theory is given by $\sum_{i=0}^{\frac{k+1}{3}}p_i$ and each phase has a Higgs branch given by an $\prod_{i=1}^l S_{n_i}$ orbifold of the finite tension theory \eref{k2finite}.
\be
\CH_{\left\{n_0; n_i\right\}}\left( ~ \node{}{0} - \node{}{9} - ~ \cdots ~ - \node{}{3k-15} - \node{\wver{}{3k+3}}{3k-6} \qquad\right) = \CC\left({\sf S}_{k,j,\left\{{n_i}\right\}}\right) = .
\ee
This analysis demonstrates the rich structure of phases of these 6 dimensional theories, where in each phase there is a different Higgs branch, many of them are non Abelian discrete orbifolds of the finite tension theory.

\section{The relationship between 6d theories and 3d mirror quivers}\label{relationship}

In this section we elaborate on the connection between the 6d SFTs studied in this paper and the 3d mirror quivers employed to study their Higgs branches. This relies on recent understandings regarding the torus compactification of various 6d SCFTs to 4d, where they are related to various class S type theories. With this knowledge, and the results of \cite{Benini:2010uu}, we can further compactify to 3d to get Lagrangian mirror duals, which are the quivers studied in the previous sections. Since the Higgs branch is invariant under dimensional reduction, the Coulomb branch of the 3d mirror should be the same as the Higgs branch of the original 6d theory. Two classes of theories are studied, and each class has slightly different details. Hence we study each in turn.  

\subsection{M5-branes on $\BC^2/\BZ_k$}

We start with the case of 6d SCFTs living on M5-branes on a $\BC^2/\BZ_k$ singularity, which are the theories of concern in Section \eref{sectionM5}. As discussed there, these can be constructed in Type IIA string theory, from which it is apparent that they posses a low-energy description as a quiver gauge theory on a generic point on their tensor branch. They are determined by the number of M5-branes $n$ and the order of the orbifold group $k$. 

\begin{figure}
\center
\includegraphics[width=1\textwidth]{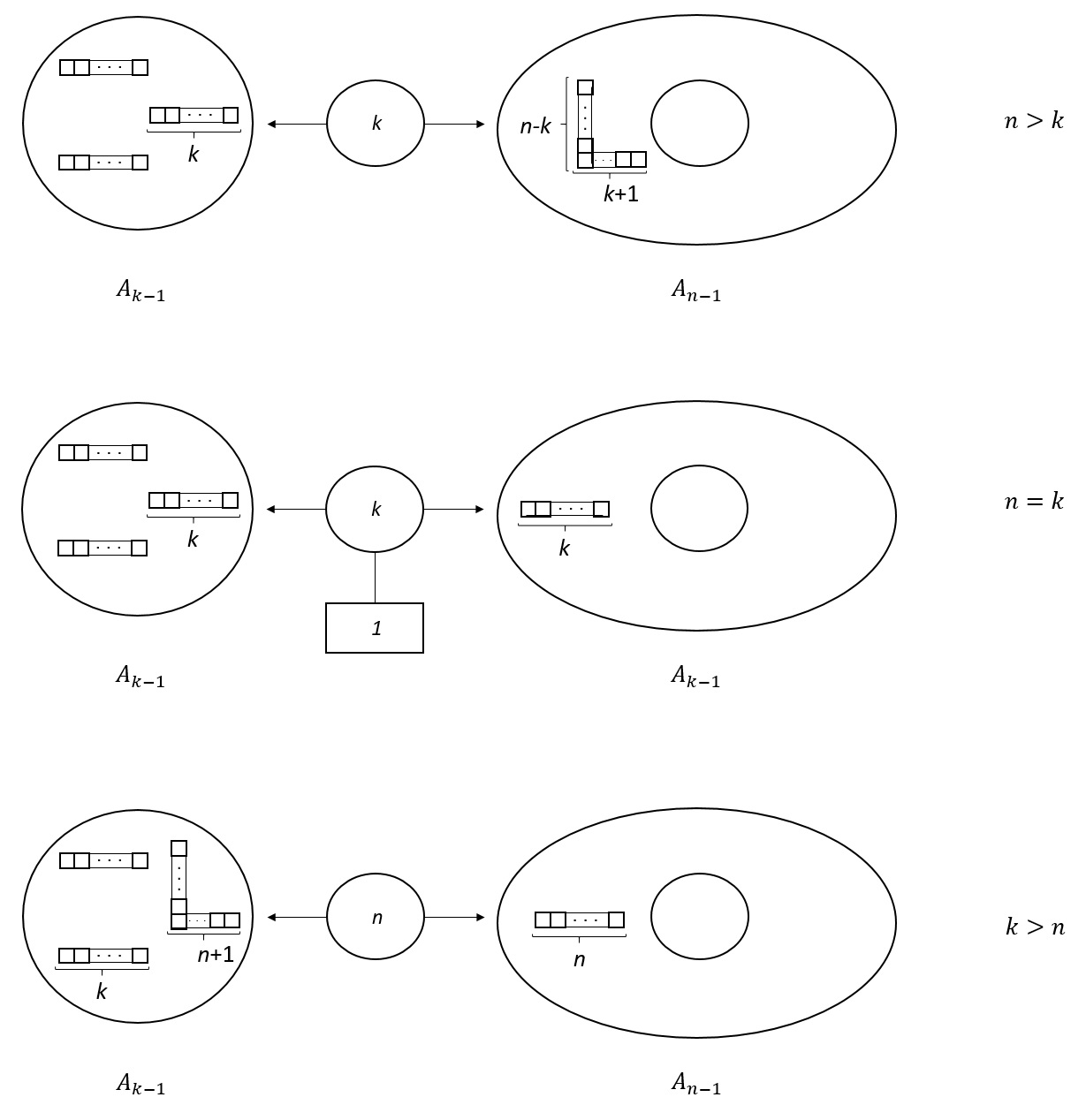} 
\caption{The 4d theories conjectured by \cite{Ohmori:2015pia} to result from the torus compactification of the 6d SCFTs living on $n$ M5-branes on a $\BC^2/\BZ_k$ singularity. The form of the theories differ depending on the relative size of $n$ and $k$. Here the pieces on the left and right correspond to two class S theories associated with a three punctured sphere and a one punctured torus respectfully, where beneath each class S theory, the type of $(2,0)$ theory of the class S theory is written. The two theories are connected through the middle gauge group, which gauges the punctures symmetries at the tip of the arrows coming out of it.}
\label{4dTheories}
\end{figure}

We next want to consider reducing these theories to 3d, and begin by first considering the reduction to 4d. This was studied in \cite{Ohmori:2015pia} who formulated a conjecture for the resulting 4d theories. These can be described in terms of class S theories, and differ slightly depending on whether $n>k$, $n=k$ or $n<k$. The explicit theories are shown in Figure \ref{4dTheories}. Note that the gauging in these theories is IR free.

We next consider the reduction to 3d. Since the gauging is IR free these should just reduce to the analogous gauging of whatever the two class S theories reduce to. While these are generally non-Lagrangian theories, in 3d they have a Lagrangian mirror as a star shaped quiver \cite{Benini:2010uu}. So the resulting theory can be described as these two star shaped quiver gauge theories connected via gauging a global symmetry which, as we have taken the mirror dual, acts on the Coulomb branch of both theories. 

\begin{figure}
\center
\includegraphics[width=1\textwidth]{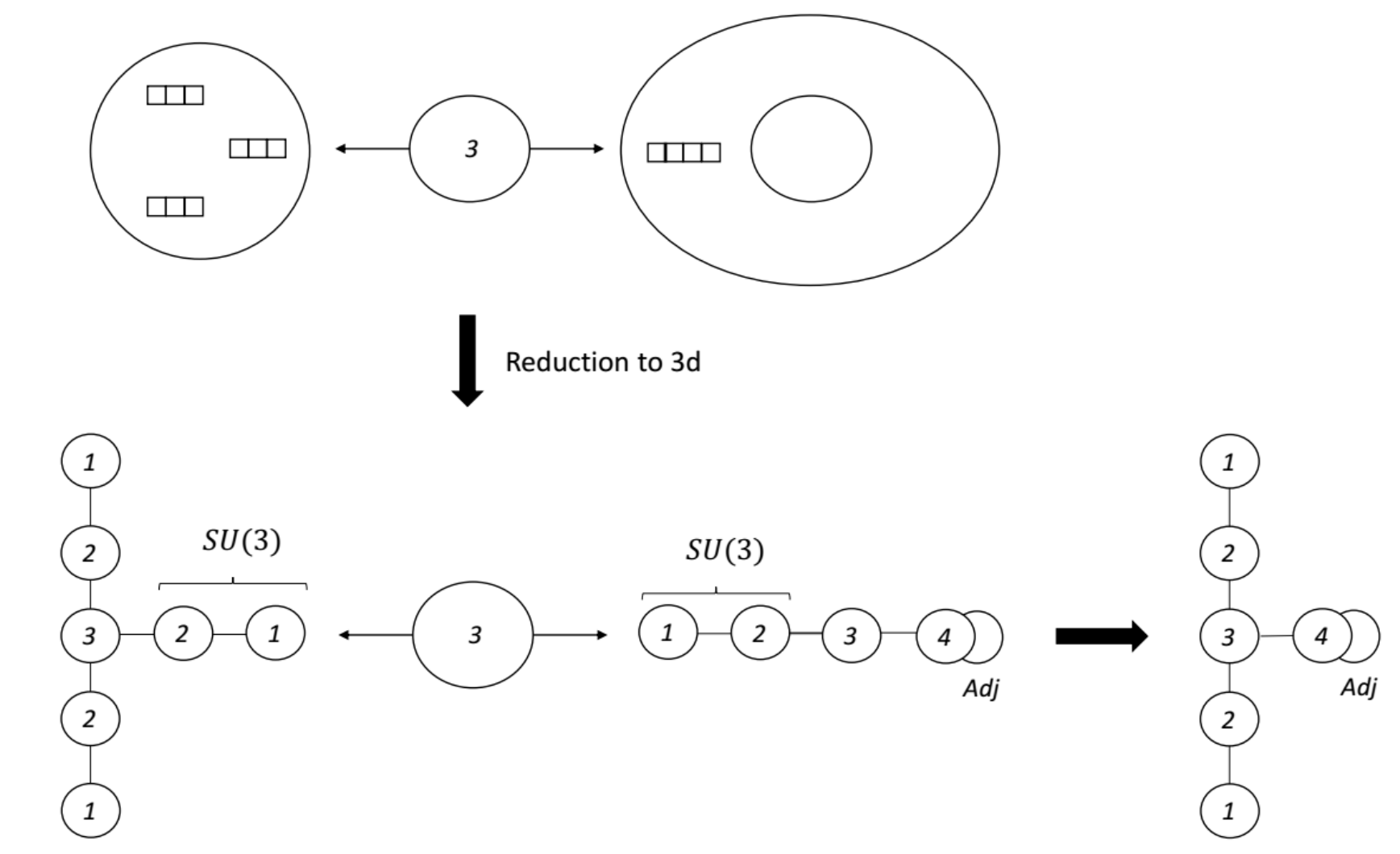} 
\caption{An example of the 3d reduction when $n=4, k=3$. First, we reduce the 4d theory to 3d. As the $SU(3)$ gauging is IR free, we can just reduce the two class S theories, which have mirror Lagrangian duals, and connect them via the same gauging. The resulting theory is shown in the bottom right, where on the left is the star shaped mirror duals of $T_3$, and in the middle is the dual of the torus class S theory, where the circular line on the last node stands for an adjoint hyper. The two theories are connected by gauging the Coulomb branch $SU(3)$ symmetry associated with the quiver tail singled by the parenthesis. This gauging leads to the  identification of the nodes connected to the quiver tails leading to the 3d quiver in the bottom right.}
\label{Example}
\end{figure}

Consider the case of $n>k$, then the $SU(k)$ that we are gauging is associated with the quiver tail of the form $U(1) \times U(2) \times ... \times U(k-1)+kF$. An important result in 3d dynamics here is that such a gauging acts as kind of a Delta function identifying the two $SU(k)$ groups. The resulting theory flows to a theory, which can be described by taking the two quivers, removing the quiver tail associated with the $SU(k)$ symmetry that was gauged and adjoining them at the node connected to the quiver tail. An example of this procedure for $n=4, k=3$ is shown in Figure \ref{Example}. Applying this procedure to general cases leads to the 3d quiver (\ref{mirrorUkinfinite}).  

Similar steps can be carried over also for the $k<n$ case and also lead to the quiver (\ref{mirrorUkinfinite}). The $n=k$ case is a bit trickier as we have the additional fundamental hyper multiplet under the $SU(k)$. This fundamental hyper can be absorbed into one of the gauged quiver tails, where it increases the rank of all the unitary groups in the tail by $1$ and also adds a $U(1)$ to the end of the quiver. For instance, consider the $k=2$ case where the quiver tail looks like $U(2) \times U(1)$ and we are gauging the $SU(2)$ symmetry associated with the topological symmetry of the $U(1)+2F$. In this case we can add the fundamental hyper by changing one of the quivers to $U(2) \times U(2) \times U(1)$ and again gauge the $SU(2)$ symmetry associated with the topological symmetry of the $U(1)+2F$ part without the free hyper. The idea here is that due to the duality discussed in \cite{KWY}, the middle $U(2)+3F$ part becomes $U(1)+3F$ and a twisted free hyper, following which the end of the quiver, which is now $U(1)+1F$ becomes just a free twisted hyper. These two twisted hypers form a doublet under the gauged $SU(2)$. This works similarly for other values of $k$. When the dust settles, we see that we still get the quiver (\ref{mirrorUkinfinite}).   

We can now straightforwardly generalize to the case where not all the M5-branes are coincident. In 6d we can describe this situation as follows. For each collection of coincident M5-branes we have a 6d SFT of the type we considered, with the number of coincident M5-branes playing the role of $n$ for that collection. These are connected to one another through gauging of their $SU(k)$ global symmetry, the gauging being IR free. Therefore, we expect the 3d reduction of this situation to be given by the 3d quiver we associate with each 6d SFT connected again via gauging of the $SU(k)$ groups. The $SU(k)$ groups appear on the Coulomb branch as the symmetries of the two quiver tails that emanate from the central node. By the previous statement, this gauging just amounts to adjoining all these 3d quivers along the central node leading to the 3d quiver in (\ref{quiverpartitions}).    

From these 6d SCFTs we can get additional 6d SCFTs by going on the Higgs branch, particularly the one associated with the mesons charged under the $SU(k)$ global symmetries. This leads to SCFTs with a low-energy description on the tensor branch as quivers of the form $SU(k_0) \times SU(k_1) \times \cdots \times SU(k) \times \cdots \times SU(k) \times \cdots \times SU(k'_1) \times SU(k'_0)$ for $k_0 \leq k_1 \leq \cdots \leq k \geq \cdots \geq k'_1 \geq k'_0$. Determining the theory then also requires specifying the quiver tail structure at the two ends. This is known to be conveniently represented by a partition of $k$ or a Young diagram with $k$ boxes as in \cite{Gaiotto:2009we}. So these more general 6d SCFTs are specified by the numbers $n, k$ and two Young diagrams with $k$ boxes. There is a set of KP transitions \cite{Cabrera:2016vvv} that connect such theories and gives the general structure under a Hasse diagram, one for each side.

It is straightforward to extend our result for the 3d quivers also to those cases. The two Young diagrams associated with the 6d SFT appear in the 4d theories of Figure \ref{4dTheories} as the Young diagrams of the class S theory on the left. The 4d reduction, then, for these more general cases is given by changing these Young diagrams. The associated 3d quivers change then by modifying the two quiver legs. 
 
In Sections \eref{subsectionSU3} and \eref{subsectionG2} we consider theories belonging to this more general family. Particularly the theories there are for $k=2$ with $n=3$ and $n=4$, respectively, and we have closed one or both of the ungauged minimal punctures in the $T_2$ theory. On the 3d quiver this has the effect of removing a $U(1)$ gauge node associated with each  closed minimal puncture. From this, for $k=2, n=3$ we indeed get one closed minimal puncture resulting in quiver \eref{31quiver}, and two closed minimal punctures for $k=2, n=4$, resulting in quiver \eref{4quiver}.

\subsection{M5-branes on $\BC^2/\BZ_k$ in the presence of an M9-plane}

The second case we consider are theories that can be constructed in string theory as those living on M5-branes on $\BC^2/\BZ_k$ in the presence of an M9-plane. Again we shall first consider the reduction to 4d on a torus, and then further reduce to 3d. Various aspects of this problem were analyzed in \cite{Ohmori:2015pi,Hayashi:2015fsa,Zafrir:2015rga,Hayashi:2015zka,Ohmori:2015p,Mekareeya:2017jgc}, and for the cases we consider here we can just use the 3d quiver prescription given in \cite{Mekareeya:2017jgc}. 

For completeness we shall briefly review the logic of the derivation. As mentioned in the previous sections, we can in many cases reduce the M-theory description to a brane configuration in Type IIA string theory. When applied to this case, this procedure results in a brane system involving D6-branes, D8-branes and NS5-branes in an O8$^-$ plane background. We can then first compactify one direction and perform T-duality. This maps the configuration to a Type IIB brane configuration containing D5-branes, D7-branes and NS5-branes in the presence of two O7$^-$ planes. The latter can be resolved to a pair of 7-branes leading to a brane web. This brane web can be used to realize various 5d gauge theories which are the low-energy limit of the compactified system in various ranges of the parameter space.

We next consider taking the zero radius limit. This can be implemented on the brane web, and leads to a web describing a 5d SCFT of the type studied in \cite{Benini:2009aa}. We can next reduce on an additional circle to 4d, which is known to give a 4d class S theory \cite{Benini:2009aa}. We can continue and reduce to 3d, where the results of \cite{Benini:2010uu} imply that we get a mirror star shaped quiver. Indeed, these are the types of 3d quivers associated with the 6d SFTs above.

In Section \eref{Smallinstantons} we use two examples of families that can both be realized as M5-branes on a $\BC^2/\BZ_k$ singularity in the presence of an M9-plane\footnote{We can also consider theories related to such a system via vevs to various fields (KP transitions), similarly to the case without the M9-plane. These can be dealt with in a similar manner to the previous case, see \cite{Mekareeya:2017jgc}.}. For these cases we can use the results of \cite{Mekareeya:2017jgc}, which give a prescription for the 3d quiver in terms of 6d data like the associated low-energy quiver semi-gauge theory on a generic point on the tensor branch. The prescription is rather involved so we refer the interested reader to the reference for details.       

The low-energy quiver gauge theories describe the 6d theory in a phase when all the M5-branes are separated. These can then be used to derive the 3d quiver, using for instance the Type IIB brane construction of the analogous 3d system as in \cite{Hanany:1996ie}. This method though can only be applied if the analogous 3d system can be constructed using a brane system.  

Finally we can consider the mixed cases, where we are not at the 6d SCFT point, yet at a special point in the tensor branch where some collection of M5-branes coincide either inside or outside of the M9-plane. We can approach this problem similarly to the way we tackle the previous case. Particularly the resulting 6d theory can be described by a series of 6d SCFTs connected via IR free gaugings of parts of their global symmetry. When reduced to 3d these should be described by an analogous theory. For the 6d SCFTs associated with collections of coincident M5-branes outside the M9-plane, we can use the 3d mirrors in the previous subsection. For the 6d SCFT associated with a collection of coincident M5-branes inside the M9-plane, we can use instead the 3d mirrors in this subsection. These are connected via gauging of the global symmetries on the Coulomb branch, which as we discussed, can be implemented by adjoining them through the node connected to the shared quiver tail, whose Coulomb branch symmetry is gauged. In this way we can generate the various quiver theories associated with these more general cases.

\acknowledgments
A.~H.~ would like to thank Noppadol Mekareeya, Ronen Plesser, Travis Maxfield, Anton Zajac, Santiago Cabrera, Rudolph Kalveks, Travis Schedler and Marcus Sperling for enlightening discussions.
G.~Z.~ would like to thank Shlomo Razamat for enlightening discussions.
A.~H.~ and G.~Z.~ gratefully acknowledge the Tsinghua Sanya International Mathematics Forum (TSIMF) for hosting the workshop on SCFTs in dimension 6 and lower where this project was initiated.
A.~H.~ would like to thank Ronen Plesser and the department of physics of Duke university for their kind hospitality during the progress of this project.
A.~H.~ would like to thank GGI for kind hospitality during the completion of this project.
A.~H.~ is supported in part by an STFC Consolidated Grant ST/J0003533/1, and an EPSRC Programme Grant EP/K034456/1.
The work of G.~Z.~ supported in part by WPI Initiative, MEXT, Japan at IPMU, the University of Tokyo.

\bibliographystyle{JHEP}
\bibliography{ref}
%\bibliography{ref,refA}

\end{document}